\documentclass[journal,twoside,web]{ieeecolor}
\usepackage{tmi}
\usepackage{cite}
\usepackage{amsmath,amssymb,amsfonts}
\usepackage{algorithmic}
\usepackage{graphicx}
\usepackage{textcomp}
\usepackage{dblfloatfix}
\usepackage{soul}
\usepackage{orcidlink}
\usepackage{multirow}
\usepackage{comment}

\def\BibTeX{{\rm B\kern-.05em{\sc i\kern-.025em b}\kern-.08em
    T\kern-.1667em\lower.7ex\hbox{E}\kern-.125emX}}
\begin{document}

\title{Characterization of Polarimetric Properties in Various Brain Tumor Types Using Wide-Field Imaging Mueller Polarimetry}

\author{Romane Gros \orcidlink{0000-0002-0911-2792}, Omar Rodríguez-Núñez \orcidlink{0000-0003-0280-9519}, Leonard Felger, Stefano Moriconi \orcidlink{0000-0002-5705-7839}, Richard McKinley \orcidlink{0000-0001-8250-6117}, Angelo Pierangelo, Tatiana Novikova \orcidlink{0000-0002-9048-9158}, \IEEEmembership{Member, IEEE}, Erik Vassella, Philippe Schucht, Ekkehard Hewer \orcidlink{0000-0002-9128-0364} and Theoni Maragkou
\thanks{Manuscript received December 12, 2023; revised February 21, 2024. This work was supported by SNF Sinergia grant CRSII5\_205904 (Polarimetric visualization of Healthy brain fiber tracts for tumor delineation during neurosurgery). Corresponding author: Romane Gros.}
\thanks{R. Gros, T. Maragkou and E. Vassella are with Institute of Tissue Medicine and Pathology, University of Bern, 3008 Bern, Switzerland (e-mail: romane.gros@unibe.ch; theoni.maragkou@unibe.ch; erik.vassella@unibe.ch). 
Additionally, R. Gros is with the Graduate School for Cellular and Biomedical Sciences, University of Bern, 3012 Bern, Switzerland.}
\thanks{O. Rodríguez-Núñez, L. Felger and P. Schucht are with the Department of Neurosurgery, Inselspital, Bern University Hospital, University of Bern, 3010 Bern, Switzerland (e-mail: omar.rodrigueznunez@insel.ch; leonardalexander.felger@insel.ch; philippe.schucht@insel.ch).}
\thanks{S. Moriconi and R. McKinley are with the Support Center for Advanced Neuroimaging (SCAN), University Institute of Diagnostic and Interventional Neuroradiology, Inselspital, Bern University Hospital, University of Bern, 3010 Bern, Switzerland (e-mail: stefano.moriconi@insel.ch; richard.mckinley@insel.ch).}
\thanks{A. Pierangelo and T. Novikova are with the LPICM, CNRS, Ecole Polytechnique, IP Paris, 91477 Palaiseau, France (e-mail: angelo.pierangelo@polytechnique.edu; tatiana.novikova@polytechnique.edu).}
\thanks{E. Hewer is with the Institute of Pathology, Lausanne University Hospital, University of Lausanne, 1011 Lausanne, Switzerland, France (e-mail: ekkehard.hewer@chuv.ch).}}

\maketitle

\begin{abstract}
Neuro-oncological surgery is the primary brain cancer treatment, yet it faces challenges with gliomas due to their invasiveness and the need to preserve neurological function.
Hence, radical resection is often unfeasible, highlighting the importance of precise tumor margin delineation to prevent neurological deficits and improve prognosis.
Imaging Mueller polarimetry, an effective modality in various organ tissues, seems a promising approach for tumor delineation in neurosurgery.
To further assess its use, we characterized the polarimetric properties by analysing 45 polarimetric measurements of 27 fresh brain tumor samples, including different tumor types with a strong focus on gliomas.
Our study integrates a wide-field imaging Mueller polarimetric system and a novel neuropathology protocol, correlating polarimetric and histological data for accurate tissue identification.
An image processing pipeline facilitated the alignment and overlay of polarimetric images and histological masks.
Variations in depolarization values were observed for grey and white matter of brain tumor tissue, while differences in linear retardance were seen only within white matter of brain tumor tissue.
Notably, we identified pronounced optical axis azimuth randomization within tumor regions.
This study lays the foundation for machine learning-based brain tumor segmentation algorithms using polarimetric data, facilitating intraoperative diagnosis and decision making.


\end{abstract}


\section{Introduction}\label{sec:introduction}

\subsection{Brain cancer and neurosurgery challenges}\label{sec:intro_generalities}
\IEEEPARstart{I}{n} 2020, approximately 300,000 cases of central nervous system (CNS) cancers were diagnosed worldwide, leading to around 250,000 deaths \cite{Sung2021}.
Neuro-oncological surgery plays a central role in the first-line treatment of brain tumors.
Gliomas, the most frequent intraaxial CNS tumors, often exhibit an infiltrative growth pattern \cite{Seker-Polat2022}.
Furthermore, they frequently develop in critical regions of the brain, known as “eloquent areas”, which house vital functions such as speech, movement, and vision.
The complete tumor resection while preserving neurological function is of utmost importance during neurosurgery, since both are closely associated with improved prognosis \cite{Rahman2017, Karschnia2022}.
Nevertheless, as healthy (HT) white matter (WM) cannot always be distinguished from tumor tissue, complete surgical resection of gliomas is often unattainable and finding a cure for these tumors remains elusive.
Hence, accurate identification of tumor boundaries as well as identification of crucial fiber tracts is critical to enhance the procedure’s effectiveness. 

Significant research efforts have been made for intraoperative techniques to distinguish healthy brain tissue (HBT) from neoplastic brain tissue (NBT).
Despite their respective merits, each one of them has drawbacks.
Intraoperative MRI depicts the extent of resection \mbox{\cite{Kubben2011}}, but is very costly, time consuming and a single-time point method \mbox{\cite{Eljamel2016, Abraham2019}}.
5-ALA based fluorescence \mbox{\cite{Alston2019, Ahrens2022}} reveals residual tumor clusters, but only in a subset of brain tumors, such as glioblastoma and CNS WHO grade 3 glioma. 
Neuronavigation provides guidance but loses accuracy during surgery due to brain shift \mbox{\cite{Gerard2021}}, and has limited value for HBT versus NBT interface identification.
Secondly, there is currently no method to identify fiber tracts in real time.
Hence, there is a significant clinical need for the development of an non-invasive, real-time, repetitive intraoperative method to reveal the HBT/NBT interface and depict fiber tracts.

\subsection{Biomedical applications of Mueller polarimetry}\label{sec:applications_IMP}

Based on these observations, we pursued the development of a new method using wide-field Imaging Mueller Polarimetry (IMP) to detect brain tumor margins intraoperatively.
Numerous studies reported the potential of polarimetry-based methods for tissue characterization, both \textit{ex vivo} \cite{Ivanov2021, Kupinski2018} and \textit{in vivo} \cite{Louie2021, Vizet2017, Qi2023}.
Mueller polarimetry has shown effectiveness in distinguishing HT from neoplastic (NP) tissues in various human organs, including the colon \cite{Ivanov2021}, the cervix \cite{Vizet2017, Kupinski2018}, the skin \cite{Louie2021}, and the larynx \cite{Qi2023}.

Mueller polarimetry is applicable in brain tissue imaging, because linear birefringence is exhibited by brain fiber tracts, the organized structures within brain WM.
Several studies reported the ability of polarimetry for the evaluation of brain fiber tract organization and their reconstruction using polarized light imaging \mbox{\cite{Axer2011, Menzel2021, Jain2021}}.
Moreover, we demonstrated in a prior study the IMP's capability to determine nerve fiber tract orientation within the imaging plane \cite{Schucht2020}.
Paired with machine learning (ML) algorithms, IMP enables the differentiation between the WM, housing these nerve fibers, and the grey matter (GM) \cite{McKinley2022}.
Additionally, our tailored wide-field IMP system exhibits resilience in simulated surgical environments, such as the presence of blood, sample tilting, and presence of complex surface topography \cite{RodriguezNunez2021, Felger2023}.
In addition, IMP offers specific advantages like near real-time imaging capabilities, non-invasiveness, and relatively simple instrumentation, allowing for easy integration into surgical workflows.
Altogether, these observations underscore its potential for practical application in clinical settings.

\subsection{Significance and related work}\label{sec:brain_tissue_properties}
Polarimetric properties, like linear birefringence, play a pivotal role in understanding the microscale structural organization and optical behavior of various biological HT and NP tissues \cite{Vitkin2015, Meglinski2021}.
Various optical properties of HBT and NBT have been characterized, in both \textit{in vivo} and \textit{ex vivo} tissue, using diverse optical techniques, such as optical coherence tomography \mbox{\cite{Strenge2022}}, hyperspectral imaging \mbox{\cite{Urbanos2021, Giannantonio2023}}, and Raman spectroscopy \mbox{\cite{Jermyn2015, Zhang2022}}.
However, there's a knowledge gap in quantifying the polarimetric properties of HBT and NBT.
Hence, characterizing these polarimetric properties is a crucial step in our approach.
Bridging this gap would provide deeper insight into brain tissue's optical response to probing polarized light and alterations induced by brain tumor growth.
The identification of potential polarimetric biomarkers enhancing image contrast between HBT and NBT, and the development of a database comprising accurately annotated polarimetric measurements, could help in the further development of imaging polarimetric tools and associated ML segmentation approaches for the intraoperative brain tumor delineation.
We opted to utilize \textit{ex vivo} brain tissue measurements since obtaining high quality \textit{in vivo} polarimetric measurements was not possible with the current version of the instrument.
Additionally, obtaining accurate ground truth for tissue identification posed challenges with \textit{in vivo} tissues.

\subsection{Summary of the contribution}\label{sec:summary_contribution}
In this study, the custom-built wide-field IMP \cite{Felger2023} was used to image brain tumor specimens immediately after their surgical removal.
The samples included a combination of both tumor center, with a high tumor cell concentration, and infiltration zone, with a lower concentration of tumor cells. 
A neuropathology protocol was designed to i) obtain the ground truth (GT) in terms of histological characterization of brain tissue (i.e., neuropathological diagnosis) and ii) correlate this diagnosis to the polarimetric properties of the sample.
A tailored image processing pipeline mapped the GT and polarimetric parameter maps.
By assessing the variations in polarimetric parameters across different brain tissue types, we were able to quantify for the first time the polarimetric properties of fresh NBT from various brain tumor types and contrast them with HBT.
Our findings highlight the IMP's potential in distinguishing HBT and NBT.
Leveraging the IMP's distinctive capabilities, we aim to contribute to the development of real-time intraoperative imaging tools, that will be of great help to the neurosurgical community by achieving more precise tumor resections.

\section{Methods}\label{sec:methods}

\begin{figure*}[t]
    \begin{center}
    \includegraphics[width=\textwidth]{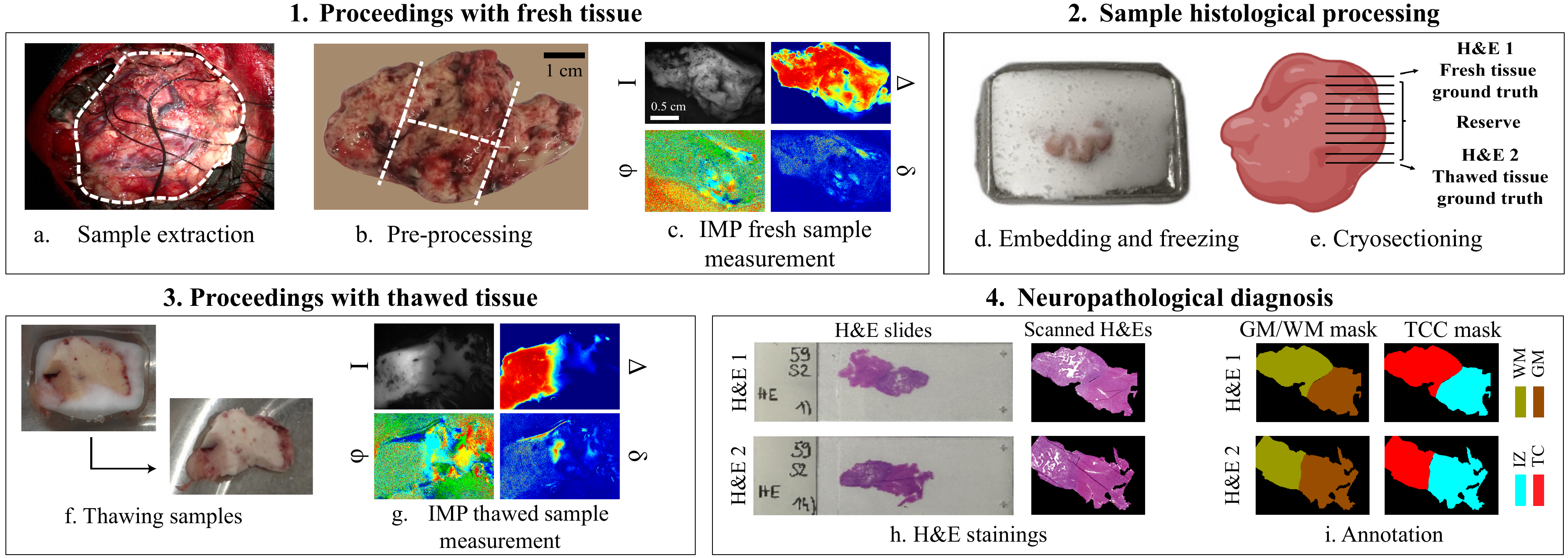}
    \end{center}
    \caption{The neuropathology protocol for correlating polarimetric parameters and GT for tissue identification in brain tumor samples: (a) brain surface image during resection with a white dashed line showing the borders of the excised tissue; (b) sample cutting into $\approx 2 \times 2$ cm sections, if needed; (c) initial acquisition of polarimetric parameters of a fresh sample; (d) Freezing at -80$^\circ$C; (e) cryosectioning; (f) sample thawing in PBS; (g) subsequent polarimetric parameter acquisition of thawed sample; (h) H\&E staining and digital scanning of the first and last sections; (i) annotation by a neuropathologist, providing GT for tissue identification.}\label{fig:protocol_description}
\end{figure*}

\subsection{Imaging Mueller polarimetry system} \label{sec:imp_system}

We examined brain tumor samples using the wide-field IMP operating in reflection configuration in a visible wavelength range, with a field of view (FOV) $2.4\times 2.1$ cm.
Detailed information on the design, calibration, and optimization of our IMP instrument has been previously reported in other studies \cite{LaudeBoulesteix2004, Lindberg2019, Kupinski2018, Schucht2020, RodriguezNunez2021, Felger2023, Gros2023}.
To summarize, the instrument described consists of an incoherent white light source, a Polarization State Generator (PSG) for modulating incident light polarization, and a detection arm with a Polarization State Analyzer (PSA), spectral filter wheel and a CCD camera for image registration.
The PSG utilizes a linear polarizer, two voltage-driven ferroelectric liquid crystals (FLC), and a further waveplate placed between the two FLCs. The PSA has the same components arranged in reverse order.
The modulation of polarization of the probing light beam is performed sequentially.
To reconstruct the Mueller matrix of a sample, the system performed 16 intensity image measurements by analyzing 4 linearly independent polarization states generated by PSG and 4 linearly independent polarization states of PSA.
To enhance the signal-to-noise ratio, we performed 8 sets of 16 intensity measurements and subsequently averaged the outcomes.
The post-processing of the recorded Mueller matrix images was done using the Lu-Chipman polar decomposition algorithm \mbox{\cite{LuChipman1996}}:
\begin{gather}
    \small
    M = M_{\Delta} M_{R} M_{D}
    \small
\end{gather}
with $M_{\Delta}$, $M_{R}$, and $M_{D}$ denoting matrices that represent three optical elements, specifically the polarizing depolarizer, the retarder, and the diattenuator.
$M_R$ represents the product of the matrix of the linear retarder $M_{RL}$ and the matrix of the circular retarder $M_{RC}$:
\begin{gather}
    \small
    M_{R} = M_{RL} M_{RC}
    \small
\end{gather}
We then extracted $M_{RL}$ using:
\begin{gather}
    \small
    M_{RL} = M_{R} M_{RC}^{-1}
    \small
\end{gather}
Our research has concentrated on investigating the total depolarization coefficient $\Delta$, the linear retardance $\delta$, the azimuth of the optical axis $\varphi$ and the greyscale intensity $I$, calculated as:
\begin{gather}
    \small
     \Delta = 1-\frac{1}{3}\vert tr(\mathbf{M}_{\Delta} - 1)\vert, \\
     \footnotesize
     \delta = \cos^{-1} \left(\sqrt{ (M_{RL_{22}}+M_{RL_{33}})^2+(M_{RL_{32}}-M_{RL_{23}})^2}-1\right), \\
     \varphi = \frac{1}{2}\tan^{{-}1} \left (\frac{M_{R_{24}}}{M_{R_{43}}}\right), \\
     I = M{_{11}}.
    \normalsize
\end{gather}
These parameters have been shown to be representative for brain tissue differentiation \cite{Schucht2020, BookBrainChapter}.  
The measurements included in this study were performed at $550$ nm. 
This wavelength might be suboptimal for certain tissue types, especially in the presence of blood due to a peak of hemoglobin absorption in the red part of the spectrum.
However, choosing $550$ nm allowed to obtain the highest quality images as the instrument currently in use was optimized for this wavelength.

\subsection{Neuropathology protocol} \label{sec:tumor_protocol}
Fresh NBT samples including both tumor center and infiltration zone were collected from the operating room immediately after their removal by a neurosurgeon (Fig. \ref{fig:protocol_description}a).
This study includes 45 measurements from 27 NBT samples containing distinguishable GM and WM, sourced from 16 patients.
The samples showcase various tumor types and subtypes (as outlined in Tab. \ref{table:overview_samples}), but were predominantly gliomas.
Informed consent was obtained from all patients.
The study received approval from the cantonal ethics committee of Bern (KEK BE 2019-02291) and adheres to the principles of the Declaration of Helsinki.

\begin{table}[h]
\centering
    \caption{Overview of the measurements included in the study.}
    \resizebox{0.9\columnwidth}{!}{

        \begin{tabular}{|c|c|c|c|}
            \hline
            Tumor type& 
            \# Measurements& 
            Mean age&
            Proportion\\
            \hline
            Glioblastoma& 
            15& 
            61.9&
            33.3\%\\
            Oligodendroglioma& 
            18& 
            40.0& 
            50.7\%\\
            Astrocytoma& 
            11& 
            40.5& 
            24.4\%\\
            Metastasis& 
            1& 
            78& 
            2.2\%\\
            \hline
            Total& 
            45& 
            52.5& 
            100\%\\
            \hline
        \end{tabular}
    }
    \label{table:overview_samples}
\end{table}

\begin{figure*}[t]
    \begin{center}
    \includegraphics[width=\textwidth]{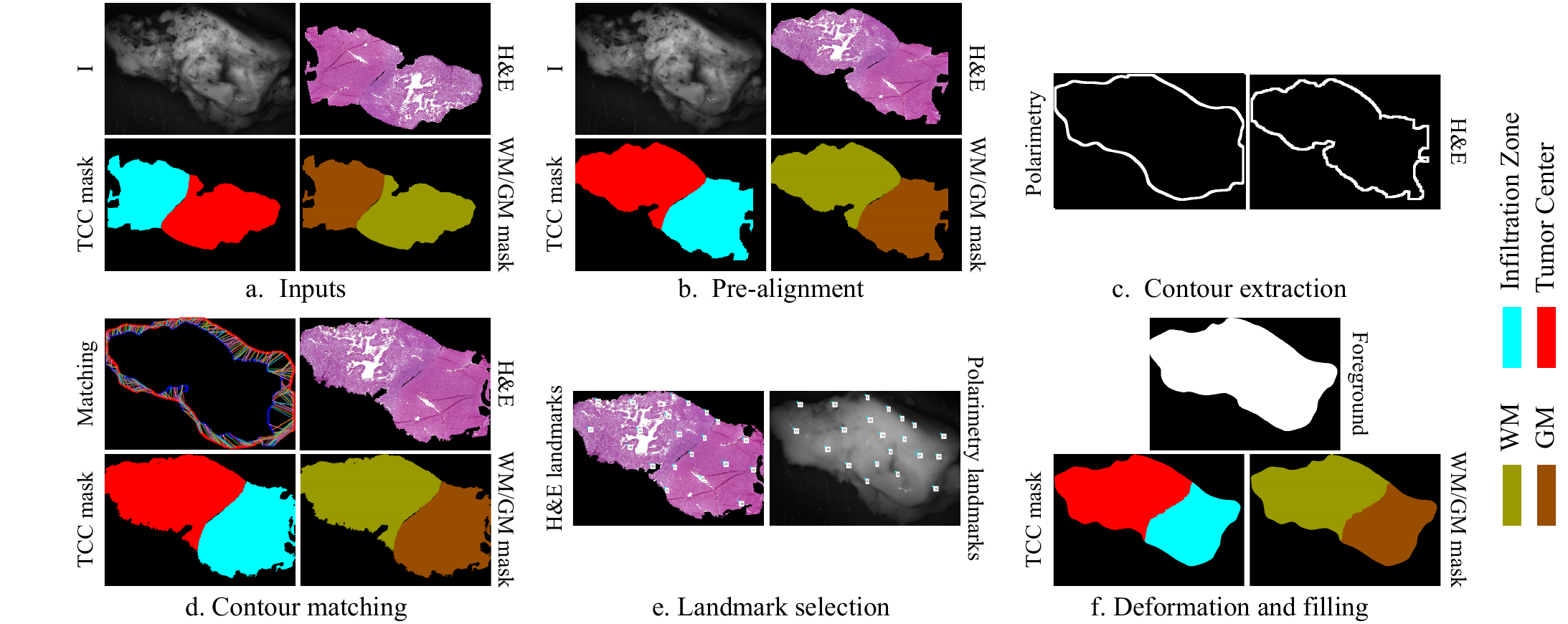}
    \end{center}
    \caption{The image processing pipeline aligning neuropathological diagnosis with polarimetric features: (a) inputs: polarimetric greyscale intensity image, histology scanned slide, and histological masks; (b) pre-alignment; (c) contour extraction of polarimetric and histological images; (d) contour matching and transformation of neuropathological masks; (e) manual selection of matched fiducial landmarks (white boxes) in both polarimetric and histological images; (f) deformation of histological image and histological masks based on fiducial landmarks and expansion of labels to match foreground boundaries.}\label{fig:matching_description}
\end{figure*}

Multiple sections were obtained to image lesions along various planes and directions, ensuring consistent image detail and resolution based on the camera's FOV (Fig. \ref{fig:protocol_description}b).
{The lateral dimensions of the sections were $\approx 2 \times 2$ cm and the thicknesses varied depending on the shape of the tissue being received from the OR, usually being $\approx 1$ cm.
The pre-processed fresh NBT samples, typically presenting a somewhat uneven surface, were then placed onto a glass Petri dish, rinsed with water, and underwent measurements using the wide-field IMP system (Fig. \ref{fig:protocol_description}c), enabling the first acquisition of sample polarimetric parameters.
The samples were then embedded in Optimal Cutting Temperature compound (OCT, Biosystems 81-0771-00) and frozen at -80$^\circ$C (Fig. \ref{fig:protocol_description}d).
Sectioning yielded 20 serial cryosections, or fewer if the tissue quantity was insufficient, from each sample (Fig. \ref{fig:protocol_description}e).
The first and last 10 $\mu$m thick sections underwent histological analysis with H\&E staining, providing a neuropathological diagnosis as close as possible to the measured tissue zone.
The remaining sections were stored in the -80$^\circ$C freezer for further molecular analyses.
The samples were then thawed by immersing them in phosphate-buffered saline (PBS) at room temperature (Fig. \ref{fig:protocol_description}f), and rinsed with water before optical measurements were carried out using the IMP instrument (Fig. \ref{fig:protocol_description}g) to capture the polarimetric properties of the thawed sample.

The H\&Es were scanned (Fig. \ref{fig:protocol_description}h) and subsequently annotated by a board-certified neuropathologist using QuPath \cite{Bankhead2017}.
These annotations served as the GT for tissue characterization (Fig. \ref{fig:protocol_description}i).
Specifically, annotations included three categories based on tumor cell content (TCC) within a sample zone: 0 - 10\%: HT tissue; 10 - 70\%: Infiltration Zone (IZ) and ; 70 - 100\%: Tumor Center (TC).
The two latter are subcategories of the NP category.
A TCC exceeding 70\% is considered sufficient for molecular analysis in diagnostic neuropathology \cite{Capper2018}.
Additionally, when histomorphologically possible, a second annotation of GM versus WM created a second mask for all studied sample images.

\subsection{Matching neuropathology and polarimetry}\label{sec:mapping_protocol}

The next step involved matching the histological labels to the polarimetric images of the same brain tissue sample.
We mapped histological labels onto the greyscale intensity image, enabling the integration of histological labels and polarimetric maps within the same referential.
Due to important differences between histological images of thin sections and greyscale intensity images of bulk brain tissue, caused by processing steps of the protocol, an image processing pipeline was necessary to recover spatial correspondence of delineated boundaries between histological and polarimetric images.
In our pipeline, polarimetry served as the "fixed" modality while histology was the "moving" modality.
Every manual interaction in the pipeline has been performed by experts or under supervision by experts.

Due to resolution imbalance, histological scans (with a resolution up to $200,000 \times 100,000$ pixels) were resampled to match the size of polarimetric images ($516 \times 388$ pixels) (Fig. \ref{fig:matching_description}a).
We then implemented a pre-alignment step using an affine transform with manually selected parameters on the resampled histological images (Fig. \ref{fig:matching_description}b).

Subsequently, a two-fold approach matched polarimetric and histological images.
In a first time, we extracted tissue contours, automatically for histological images using the labels and manually for the intensity images of fresh and thawed tissue (Fig. \ref{fig:matching_description}c). 
More specifically, we manually drew masks of the contours of the tissue visible in the histological image using the greyscale intensity image.}
A method described in\mbox{\cite{Kamani2016, Kamani2018}} leveraged contextual information from skeleton structures for extracting pairs of matching contour points (Fig. \mbox{\ref{fig:matching_description}}d), that were then used to align external parts of the images using a framework described in\mbox{\cite{Schaefer2006}}.

In a second time and to ensure accurate matching of internal regions, we used a step matching fiducial landmarks in both polarimetric and histological images.
A set of $N$ manually selected corresponding points was used, with $N$ varying between 15 and 35 based on the number of landmarks that could be found between image pairs (Fig. \ref{fig:matching_description}e).
The number of points was dependent on the number of matching landmarks points that could be found between the polarimetric and histological images.
Fifteen points were needed to ensure a consistent mapping of the border.
These matched points constrained and regularised the elastic deformation that would be applied to the histological images\mbox{\cite{Arganda2006}} (Fig. \mbox{\ref{fig:matching_description}}f).
Polarimetric images foreground borders, extracted from the manually drawn contour mask, are shown on all polarimetric maps in the results.
Pixels in the polarimetric foreground lacking a histological label were assigned the label of the nearest labeled pixel.

The skeleton matching was performed using an available implementation \footnote{\href{https://github.com/mmkamani7/SkeletonMatching}{https://github.com/mmkamani7/SkeletonMatching}} with the default parameters, while the images were aligned using an implementation of the moving least squares method \footnote{\href{https://github.com/Jarvis73/Moving-Least-Squares}{https://github.com/Jarvis73/Moving-Least-Squares}}. The deformation based on the landmark points was performed using the bUnwarpJ ImageJ plugin \footnote{\href{https://imagej.net/plugins/bunwarpj/}{https://imagej.net/plugins/bunwarpj/}} with the default parameters, except for $initial\_deformation$ being set to Coarse, $landmark\_weight$ to 3, and $stop\_threshold$ to 0.05.

We sought to validate the pipeline aligning polarimetric and histological images to ensure accurate matching of borders between zones of different tissue types (i.e., 70-100\% TCC vs. 10-70\% TCC or GM vs. WM) corresponding to a priori expectations.
To achieve this, we manually created a GT mask for 7 polarimetric intensity images, delineating GM and WM zones using corresponding measured greyscale intensity images.
Next, we extracted the GM/WM border by identifying pixels labeled as GM or WM with neighbors labeled differently.
We dilated the border line (BL) with a $7\times7$ unit matrix kernel through 5 iterations, resulting in a 31-pixel BL.
The dilation was applied in such a way to fit the uncertainty region, defined as the area where pixels cannot be unambiguously attributed to either GM or WM zones, with a width of approximately $1.60$ mm \cite{Gros2023}.
With each pixel representing $0.05mm$, the dilated BL corresponds to $1.55 mm$, encompassing the size of the uncertainty region.
Subsequently, labels for GM and WM zones were obtained after each step of the matching pipeline.
Similarly, we extracted the GM/WM borders and overlaid the two GT border with the aligned border to visually assess their similarity.
The dice scores were computed using the following formula:
\begin{gather}
    \resizebox{\hsize}{!}{
        $DSC = \frac{2 \times \#\:correctly\:aligned\:pixels}{\#\:BL\:pixels\:aligned\:image +\: \#\:BL\:pixels\:GT\:image}$
    }
\end{gather}
They provided an objective measurement of the similarity between the two borders.

\subsection{Quantification of polarimetric parameters}\label{sec:quantitative_combined}

We characterized the polarimetric parameters of fresh HBT based on nine measurements from fresh brain autopsy samples.
The same neuropathology protocol was applied to both NBT and HBT samples.
Masks delineating GM and WM, as well as the background (BG), with no tissue, were manually drawn for HBT using greyscale intensity image whereas for NBT, masks were obtained from the protocol described in Sections \ref{sec:tumor_protocol} and \ref{sec:mapping_protocol}.

Polarimetric parameter distributions were extracted for each combination of GM/WM and TCC regions (e.g., regions labeled as (HT+WM), or as (IZ+GM), or as (TC+WM), etc.).
The previously generated masks were used to extract depolarization and linear retardance values.
We quantified the variability in the azimuth of the optical axis at a local scale and introduced a new metric, the "azimuth local variability," by computing the circular standard deviation within 4$\times$4 pixel patches fully within the considered GM/WM and TCC regions.
We conducted computations using various sizes, ranging from 3$\times$3 pixel to 10$\times$10 pixels and found that the most pronounced contrast between NP and HT tissue was achieved by using a 4$\times$4 pixel regions.
Moreover, using larger patches leads to border blurring between the different tissue types.
Additionally, we estimated the BG noise by extracting such patches exclusively from the BG zones.
The distributions obtained from individual sample measurements across the entire sample set were combined to establish global parameter distributions in multiple tissue regions.
We normalized the distribution of the three polarimetric parameters ($\Delta$, $\delta$, $\varphi$) based on the extracted regions, ensuring that the normalization process takes into account the specific regions of interest extracted from the grayscale intensity image.
Histograms of these parameters were generated to facilitate comparisons across different tissue types and statistical descriptors, including mean, standard deviation, and median values, were computed for the distributions of depolarization, linear retardance, and azimuth local variability in different GM/WM and TCC regions.

We fitted histograms with a sum of two Gaussian distributions for the azimuth local variability in WM brain tissue, 
Two components were selected as two peaks were visible in the azimuth local variability distributions in NBT regions, and as from theory we expect to observe two kinds of tissues: one highly affected by tumor cell growth showing high azimuth local variability and one slightly affected, showing low azimuth local variability.
We then compared the parameters of two Gaussians obtained in the different TCC regions to gain insights on mechanisms governing the azimuth variability among different TCC regions.

\setcounter{figure}{2}
\begin{figure}[b!]
    \centering\includegraphics[width=\columnwidth]{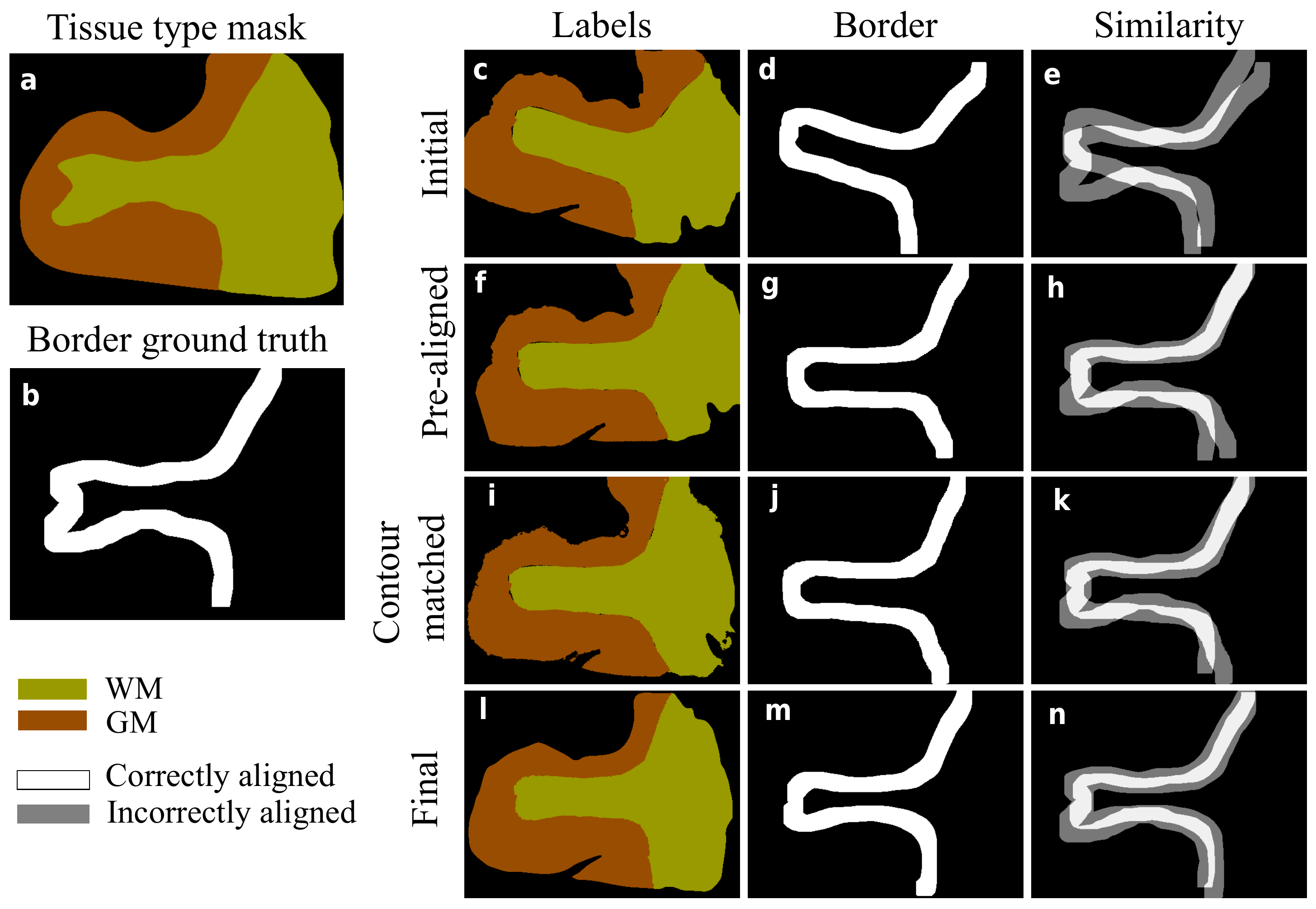}
    \caption{Matching protocol results: (a) the manually drawn tissue mask was used to extract (b) the GM/WM border. The labels (1\textsuperscript{st} column), border (2\textsuperscript{nd} column) and the over-imposition of the border with the GT (3\textsuperscript{rd} column) were obtained after four steps of the matching protocol: (c-e) the inputs; (f-h) after the pre-alignment; (i-k) after the contour matching; and (l-n) the final labels after the deformation and filling step.}
    \label{fig:results_verif_mapping}
\end{figure}

\section{Results}\label{sec:results}

\subsection{Mapping protocol}

To evaluate the similarity between the GT and the histological labels generated by the image processing matching protocol, we represented the histological labels (1\textsuperscript{st} column), GM/WM border (2\textsuperscript{nd} column), and GT-histological mask border overlay (3\textsuperscript{rd} column) after the four matching steps (Fig. \ref{fig:results_verif_mapping}).

The GT tissue mask (Fig. \mbox{\ref{fig:results_verif_mapping}}a) was used to extract the GT GM/WM border (Fig. \mbox{\ref{fig:results_verif_mapping}}b).
The 1\textsuperscript{st} row shows the input labels (Fig. \ref{fig:results_verif_mapping}c). Border from the input labels (Fig. \ref{fig:results_verif_mapping}d) misaligns with GT (Fig. \ref{fig:results_verif_mapping}e).
The 2\textsuperscript{nd} row, "Pre-aligned," represents labels after the pre-alignment step of the protocol (Fig. \ref{fig:results_verif_mapping}f). The border (Fig. \ref{fig:results_verif_mapping}g) is closer to GT, yet differences persist, especially in the lower part of the image (Fig. \ref{fig:results_verif_mapping}h).
The 3\textsuperscript{rd} row, "Contour matched," shows labels after contour matching (Fig. \ref{fig:results_verif_mapping}i). The border (Fig. \ref{fig:results_verif_mapping}j) and similarity with GT resemble "Pre-aligned" ones (Fig. \ref{fig:results_verif_mapping}k).
This was expected as the contour matching step primarily focuses on the portions of the images located close to the border with the BG.
The 4\textsuperscript{th} row represents the labels after deformation and filling, corresponding to final labels (Fig. \ref{fig:results_verif_mapping}l).
The border's shape (Fig. \ref{fig:results_verif_mapping}m) closely matches GT (Fig. \ref{fig:results_verif_mapping}n).
These observations are further substantiated by the Dice scores (Table \ref{table:dice_scores}), which quantify the similarity between the two sets of boundaries.

\begin{table}[h!]
\centering
    \caption{Dice scores for the measurement of the similarity.}
    \resizebox{0.8\columnwidth}{!}{
        \begin{tabular}{|c|c|c|c|}
            \hline
             Inputs  & Pre-alignment & \begin{tabular}[c]{@{}c@{}}Contour \\ matched \end{tabular} & Final labels \\ \hline
             0.37$\pm$0.09 & 0.63 $\pm$ 0.11    & 0.64 $\pm$ 0.09  & 0.72 $\pm$ 0.09 \\ \hline
        \end{tabular}
    }
    \label{table:dice_scores}
\end{table}

\subsection{Qualitative results}

The greyscale intensity image, alongside depolarization, linear retardance, and azimuth of the optical axis maps, is presented with aligned TCC and GM/WM masks for one HBT sample (Fig. \ref{fig:HT_sample_example}) and four NBT samples, each categorized as a different brain tumor type (Fig. \ref{fig:combined_results_tumor_90}).
All four samples contain a mixture of GM and WM tissue with regions labeled with varying TCC.
In Fig. \ref{fig:HT_sample_example}, typical HBT polarimetric parameters are visualized, showing a strong contrast between GM and WM in depolarization and linear retardance maps, as well as clearly oriented brain fibers in the azimuth of the optical axis ma.

\subsubsection{Glioblastoma}
The first sample (Fig. \ref{fig:combined_results_tumor_90}a) is diagnosed as a glioblastoma, IDH-wildtype, CNS WHO grade 4.
The presence of blood is visible on the top-left zone of the greyscale intensity image.
Depolarization values resemble those typically observed in HBT, showing a contrast, although slightly reduced, between GM and WM.
However, linear retardance values in the WM appear slightly reduced compared to the typical HBT values ($\approx20^\circ$).
Still, a small contrast persists, with higher linear retardance values in the region labeled as WM compared to the GM region.
The azimuth $\varphi$ in the (TC+WM) region exhibits considerable variability and fails to clearly visualize brain fiber tracts.
Some areas in the (IZ+GM) zone show lower azimuth variability and higher linear retardance values.

\setcounter{figure}{3}
\begin{figure}[t!]
\centering\includegraphics[width=\columnwidth]{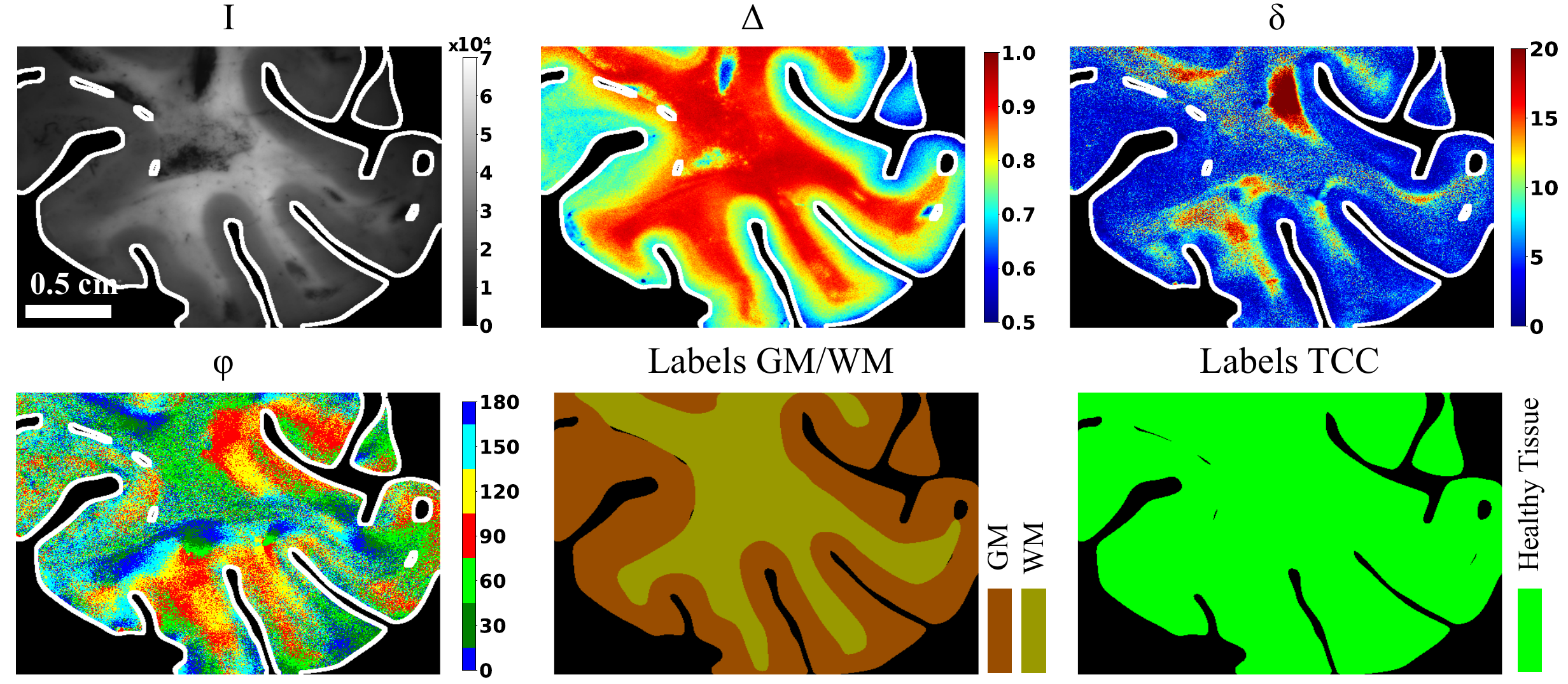}
    \caption{Polarimetric maps and histological masks of a human fresh autopsy brain sample. From top left to bottom right are represented: the greyscale intensity image $I$, the maps of the depolarization $\Delta$, the linear retardance $\delta$ and the azimuth of the optical axis $\varphi$, as well as the histological labels. The scale provided for the top left image is the same for all images.
    }
    \label{fig:HT_sample_example}
\end{figure}

\subsubsection{Astrocytoma}
The second sample (Fig. \ref{fig:combined_results_tumor_90}b) is diagnosed as an astrocytoma, IDH-mutant, CNS WHO grade 4.
Blood presence (black spots) can be observed again on the left of the greyscale intensity image.
Depolarization values resemble those in HBT samples, featuring a contrast between GM and WM, except in the region with blood, where depolarization values drop due to the peak of hemoglobin absorption around $550nm$.
Linear retardance values, however, are lower than expected in the WM region for HBT.
Once again, a pattern of noisy azimuth of the optical axis is evident in the (TC+WM) region.
Some regions in the GM exhibit low azimuth local variability and high linear retardance.

\setcounter{figure}{4}
\begin{figure*}[t!]
    \centering\includegraphics[width=0.96\textwidth]{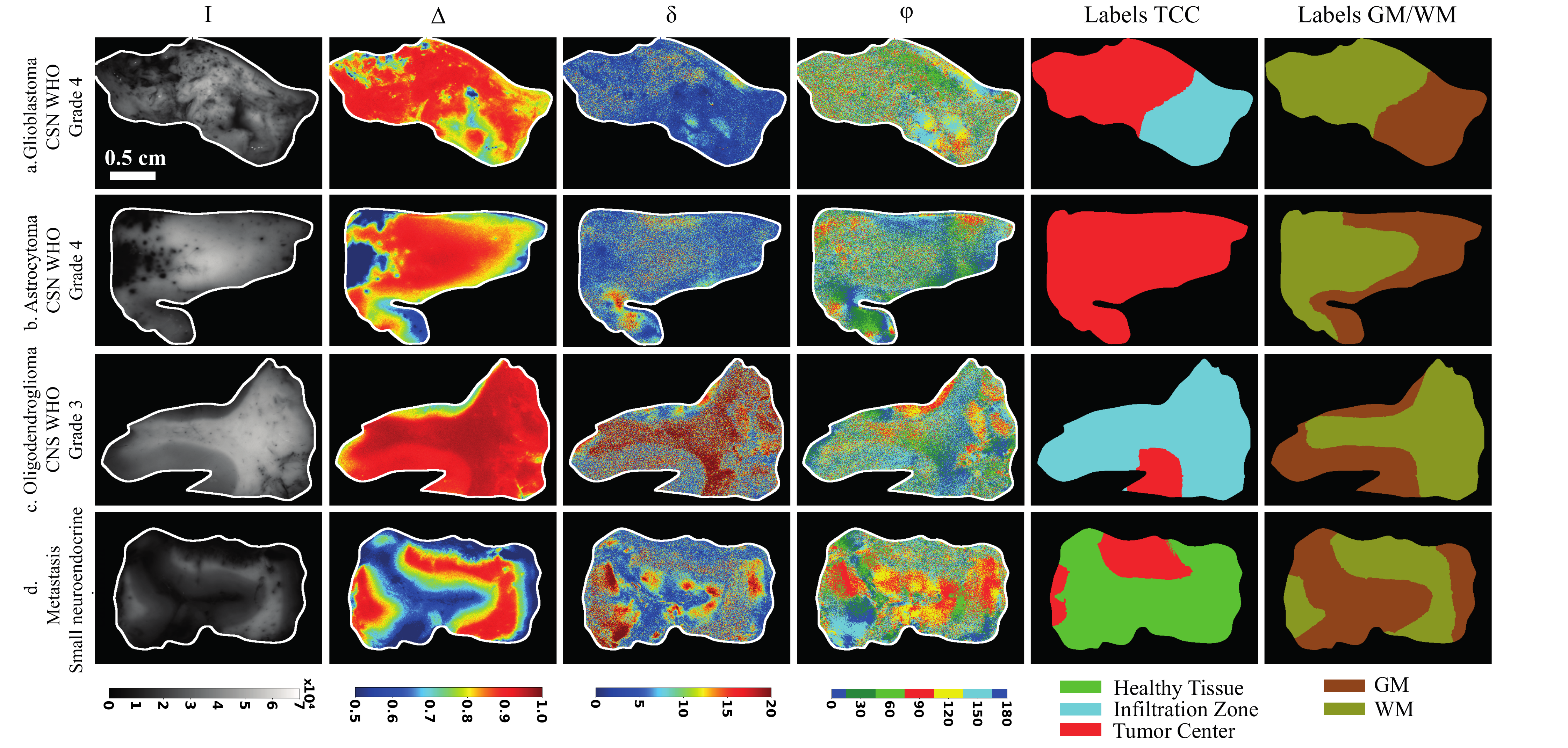}
    \caption{Polarimetric maps and neuropathological diagnosis for four different brain tumor samples: (a) a glioblastoma, IDH-wildtype, CNS WHO grade 4; (b) an astrocytoma, IDH-mutant, CNS WHO grade 4; (c) an oligodendroglioma, IDH-mutant and 1p/19q-codeleted, CNS WHO grade 3 and (d) a brain metastasis from a small cell neuroendocrine lung cancer. The columns from left to right show the greyscale intensity image $I$, the maps of the depolarization $\Delta$, the linear retardance $\delta$ and the azimuth of the optical axis $\varphi$, as well as the labels for TCC and GM/WM masks. The scale provided for the top left image is the same for all images.}
    \label{fig:combined_results_tumor_90}
\end{figure*}

\subsubsection{Oligodendroglioma}
The third sample (Fig. \ref{fig:combined_results_tumor_90}c) represents an oligodendroglioma, IDH-mutant and 1p/19q-codeleted, CNS WHO grade 3.
Visual inspection of the depolarization and linear retardance maps reveals that both parameters display properties similar to those in HBT, such as higher depolarization and linear retardance values in the WM zone compared to the GM zone.
Similar to the first sample, the azimuth $\varphi$ exhibits important variability in the (IZ+WM) region.
The variability however appears less important than in the first two samples.
This sample differs from the first two samples as the region labeled as WM is also labeled as IZ and not TC, which might explain the different results.

\subsubsection{Brain metastasis}
The fourth sample (Fig. \ref{fig:combined_results_tumor_90}d) is a brain metastasis from a small cell neuroendocrine lung cancer.
There is an important amount of blood in the top part of the sample and in the gyrus located at the center of the greyscale intensity image.
The depolarization values for both GM and WM zones are similar to those in HBT, except in the regions with blood.
Regarding the linear retardance map, we observe high values of retardanc in some regions of GM, while the values in the WM zone are lower than the typical values for (HT+WM) regions, although slightly higher than the values observed in the (TC+WM) zones in the first and second samples.
Similar to the other samples, the same pattern of noisy azimuth $\varphi$ is found in the (TC+WM) region, with a lower azimuth $\varphi$ variability pattern in the (HT+WM) region.
We also observe some regions in the GM exhibiting quite low azimuth local variability and high linear retardance. 

\setcounter{figure}{5}
\begin{figure}[t!]
\centering\includegraphics[width=\columnwidth]{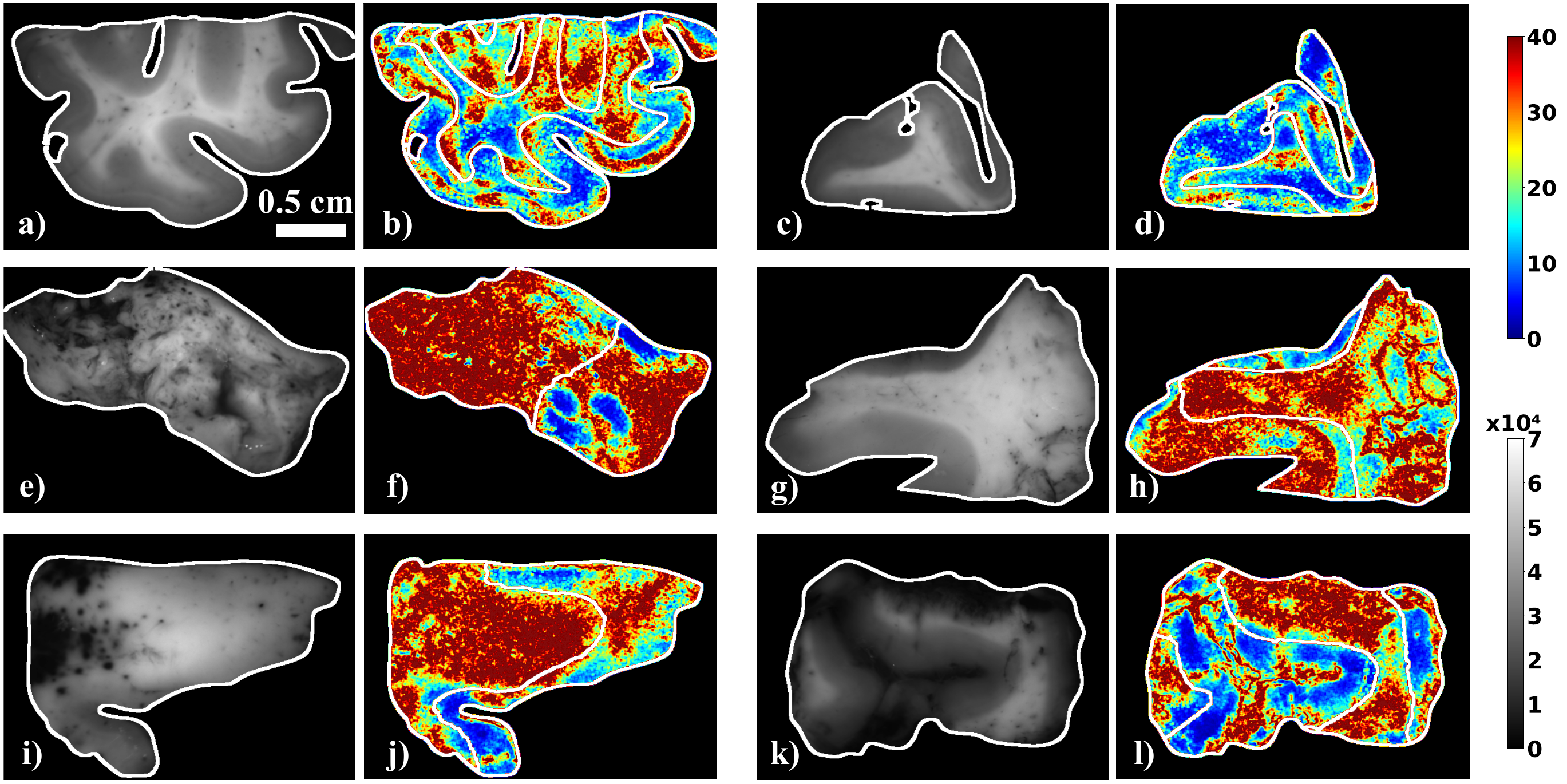}
    \caption{Greyscale intensity images and corresponding maps of the azimuth local variability for (a-d) two HT human brain specimens and (e-l) four analyzed tumor samples from Fig. \ref{fig:combined_results_tumor_90}. The scale provided for the top left image is the same for all images.}
    \label{fig:results_azimuth_noise}
\end{figure}

\subsubsection{Characterization of the azimuth local variability}

To better comprehend the extent of azimuth randomization within (NP+WM) areas compared to (HT+WM) zones, we implemented a visualization method for the azimuth local variability (Fig. \ref{fig:results_azimuth_noise}).
This method involved calculations of the circular standard deviation for $4\times4$ pixel patches, which served as a metric for quantifying the local variability level for the azimuth $\varphi$.
We plotted the images of human HBT obtained from a brain autopsy to establish a reference for the typical azimuth local variability level in HBT (Fig. \ref{fig:results_azimuth_noise}a-d).
While some localized WM subregions exhibit high variability level in HBT, the variability remains low in the WM, typically below $10^\circ$.
This opposes to the azimuth local variability level observed in the NBT (Fig. \ref{fig:results_azimuth_noise}e-l), where it is elevated in the WM region compared to human HBT, exceeding $35^\circ$.
Conversely, we did not observe any increase in variability level in the GM areas, except for the glioblastoma sample (Fig. \ref{fig:results_azimuth_noise}e, f).

In summary, we found that the values of depolarization in the GM ($\Delta\leq0.87$) and WM ($\Delta\geq0.87$) of NBT are similar to those of HBT.
However, this is not the case for linear retardance, as the $\delta$ values for (NP+WM) zones are lower than those in the (HT+WM) regions for three out of the four tumor samples.
We also observed a pattern of high local variability observed for the azimuth of the optical axis in all four NBT samples, particularly, in the regions labeled as WM.

\setcounter{figure}{6}
\begin{figure*}[t!]
\centering\includegraphics[width=\textwidth]{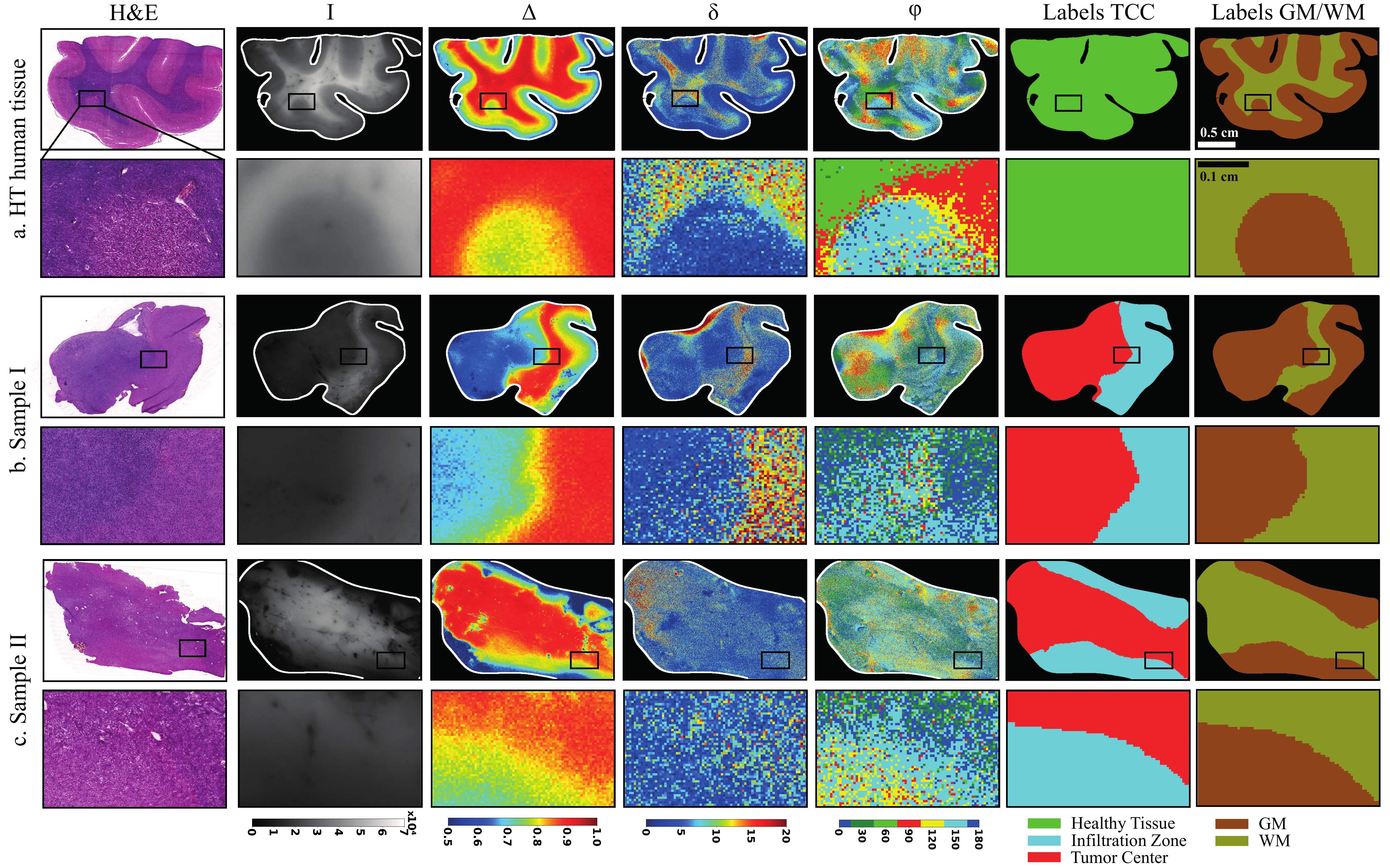}
\caption{Close-up view of the border region between GM and WM, along with various TCC zones: (a) HT human brain sample; (b, c) two NP human glioblastoma samples. For each sample the bottom row provides an enlarged view of the region indicated by a black box in the top row image. From left to right: H\&E images, greyscale intensity image, maps of the depolarization, linear retardance, and azimuth of the optical axis, TCC labels and GM/WM labels, are displayed for each sample. The scale is the same for all images and is indicated in the top right images.}
\label{fig:zoom-up-combined_together}
\end{figure*}

\subsection{Investigation of the marginal zones}

To assess the local impact of tumor on polarimetric parameters, especially at tissue border zones, we focused on these regions and analyzed the polarimetric parameters of two NBT samples and one HBT sample (Fig. \ref{fig:zoom-up-combined_together}a).
The two NBT samples are glioblastomas, IDH-wildtype, CNS WHO grade 4, featuring varying TCC regions.
For each sample and in each column, the bottom row provides a zoomed view of the region indicated by a black rectangle in the top row image.

\subsubsection{HT tissue}
Polarimetric parameters are visualized at the GM/WM border for HBT (Fig. \ref{fig:zoom-up-combined_together}a), enabling a comparison between HBT and NBT.
Depolarization maps accurately delineate the border, with higher values in WM than in GM.
This trend remains in linear retardance maps, offering a clear contrast between tissue types.
The azimuth of the optical axis map displays low local variability, highlighting a well-defined U-fiber represented in green and red/yellow, from left to right.

\subsubsection{Glioblastoma sample I}
In the first NBT sample, we examined the region at the GM/WM border, also containing the border between the IZ and the TC (Fig. \ref{fig:zoom-up-combined_together}b).
The H\&E image shows a contrast between the TC region on the left and the IZ region on the right, with higher cellularity in the former than the latter.
The depolarization map indicates comparable values to HBT in the (TC+WM) region, and the GM/WM border is clearly delineated, with higher values in WM than in GM regions.
This trend is consistent with the linear retardance maps, with a clear delineation between the tissue types and a contrast similar to the one on HBT.
However, the azimuth of the optical axis map exhibits substantial variability compared to HBT, with no clearly defined fiber in WM being visible in the zoomed azimuth map.

\subsubsection{Glioblastoma sample II}
We also explored the border between GM and WM in a second NP sample (Fig. \ref{fig:zoom-up-combined_together}c), also containing the border between the IZ and the TC.
The H\&E image reveals a contrast between GM and WM, with cellular bodies present in the GM region and absent in the WM region.
Cellularity in this sample appears lower than in Sample I.
Similar to the first sample, the GM/WM border is distinctly visible in both the greyscale intensity image and the depolarization map.
Linear retardance values in the (TC+WM) region are slightly higher than those in the (IZ+GM) region, however, they are lower than in other NBT and HBT samples.
A remarkably high variability level is observed in the azimuth of the optical axis map.

In summary, our observations suggest that tumor development does not importantly alter depolarization maps, maintaining a clearly defined GM/WM border.
Nevertheless, the contrast in linear retardance maps is greatly reduced in Sample II.
NBT exhibit elevated azimuth local variability in WM regions labeled as TC compared to HBT.

\setcounter{figure}{7}
\begin{figure*}[t!]
\centering\includegraphics[width=\textwidth]{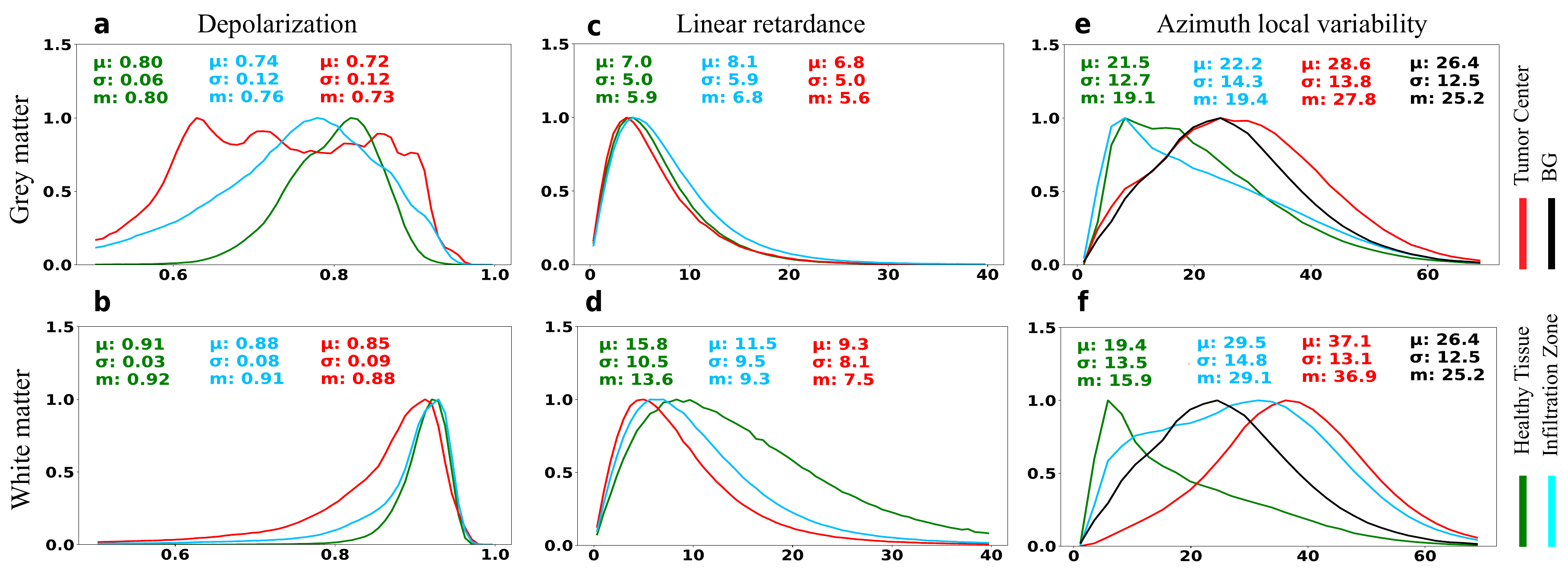}
\caption{
Normalized distributions of the polarimetric parameters in GM (first row) and WM (second row): (a, b) depolarization; (c, d), linear retardance; (e, f), azimuth local variability for HBT, NBT and BG zone.}
\label{fig:quantiative_tumor_results}
\end{figure*}

\subsection{Quantitative analysis of polarimetric parameters}

We collected data on depolarization (Fig. \ref{fig:quantiative_tumor_results}a, b), linear retardance (Fig. \ref{fig:quantiative_tumor_results}c, d), and azimuth local variability (Fig. \ref{fig:quantiative_tumor_results}e, f; see Sec. \ref{sec:quantitative_combined} for more details) for both HBT and NBT with varying TCC.
Descriptive statistics, including means ($\mu$), standard deviations ($\sigma$), and medians ($m$), were calculated for all parameter distributions.

\subsubsection{Depolarization}
Distinct depolarization value distributions in GM tissue (Fig. \mbox{\ref{fig:quantiative_tumor_results}}a) are observed for HBT and NBT.
Median depolarization values in (NP+GM) tissue consistently remain lower than those in (HT+GM) tissue.
A compelling trend emerges as TCC decreases, with depolarization distributions gradually becoming more akin to (HT+GM) tissue.
Notably, (IZ+GM) tissue shows the most resemblance, while (TC+GM) tissue displays the greatest disparity.

A similar trend is observed for depolarization values within WM (Fig. \ref{fig:quantiative_tumor_results}b), with smaller changes in distribution parameters as TCC decreases.
The distribution for (IZ+WM) tissue exhibits the highest similarity to (HT+WM) tissue, while the (TC+WM) region remains the most dissimilar.
Certain regions in NP tissues show lower depolarization values ($<$0.8), a feature virtually absent in HBT,
However, the differences in median depolarization values are relatively small, with the maximal difference being 4\% between the medians for the (HT+WM) and (TC+WM) regions.

\subsubsection{Linear retardance} 
In GM (Fig. \ref{fig:quantiative_tumor_results}c), NBT exhibits highly similar distributions for linear retardance compared to HBT.
The only noticeable difference is visible for the distributions in IZ tissue.
However, the differences in median values are minimal ($\approx1^\circ$), falling within the measurement error range of the instrument ($\approx 3^\circ- 4^\circ$).

Linear retardance values in (NP+WM) tissue generally appear smaller compared to those in (HT+WM) tissue (Fig. \ref{fig:quantiative_tumor_results}d), aligning with the previous visual observations in the assessed samples.
It is important to note that the discrepancies in median values across different TCC regions are relatively small, reaching a maximum difference of $1.1^\circ$ between the median values for IZ and TC tissue, compared to a more substantial difference of $3.4^\circ$ between the IZ and HT tissue.

\begin{figure*}[t!]
\centering\includegraphics[width=\textwidth]{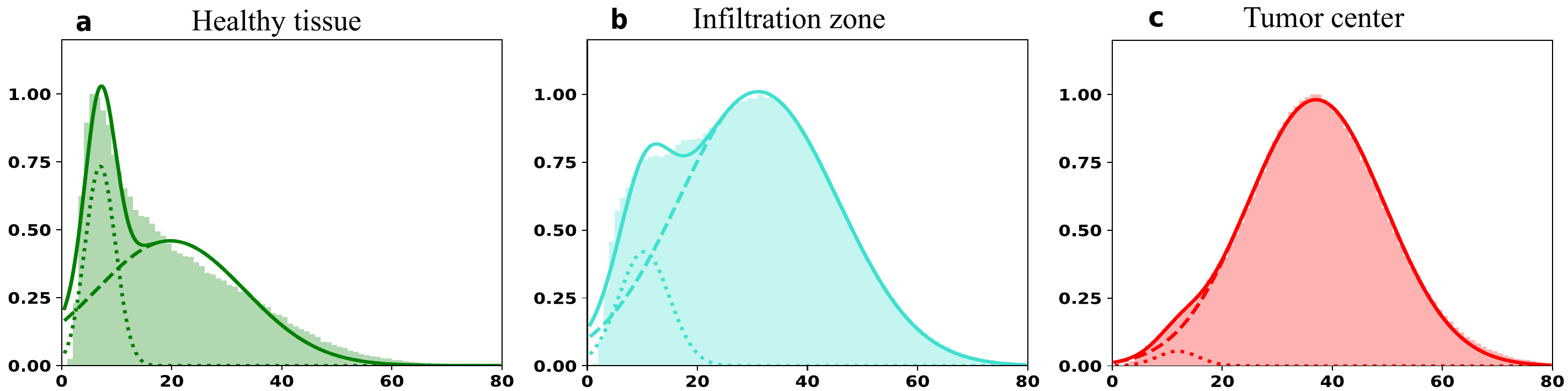}
\caption{Two-Gaussian fit for the normalized distributions of azimuth local variability in WM tissue zones: (a) (HT+WM), (b) (IZ+WM) and (c) (TC+WM).}
\label{fig:gaussian_fits}
\end{figure*}

\subsubsection{Azimuth local variability}
Both (HT+GM) and (IZ+GM) tissues exhibit similar azimuth local variability distributions (Fig. \ref{fig:quantiative_tumor_results}e), with values on average lower than the BG noise.
These distributions typically range between $10-20^\circ$, while the BG noise peaks at approximately $26^\circ$.
The azimuth local variability values in the (TC+GM) region were notably higher than those in the (HT+GM) and (IZ+GM) regions, with a distribution similar to the one for the BG.
The variations in azimuth local variability values within different regions could be partly attributed to the absence of oriented structures in GM determined by anatomical factors and spatial resolution of the IMP system.

The distribution for (HT+WM) tissue showcases a peak around $10^\circ$ (Fig. \ref{fig:quantiative_tumor_results}f), likely corresponding to the presence of oriented neurological fibers.
The peak located around $10^\circ$ is also present in the distribution for (IZ+WM) tissue, but it appears that more regions exhibit a higher level of variability compared to the corresponding distribution for (HT+WM) tissue.
The distribution of azimuth local variability for the (TC+WM) tissue demonstrates a noticeable peak around $35^\circ$, indicating an important level of azimuth local variability in NBT.
Importantly, the pattern of the azimuth local variability for (TC+WM) tissue is different from the BG noise. Indeed, the peak present in the distribution for (TC+WM) tissue is located around $40^\circ$, and the distribution is shifted towards higher azimuth local variability values. The difference in the distributions, therefore, suggests that the high azimuth local variability values observed in (TC+WM) tissue are linked to NBT properties.
The IZ could therefore be considered a transition region between the clearly defined fibers in (HT+WM) tissue and the disordered tissue organization in the (TC+WM) region.

To further gain insights into the mechanism of azimuth randomization, we performed the two-Gaussian fit of the azimuth local variability distributions in (HT+WM) (Fig. \ref{fig:gaussian_fits}a), 
(IZ+WM) (Fig. \ref{fig:gaussian_fits}b) and (TC+WM) tissue (Fig. \ref{fig:gaussian_fits}c).
The values of the parameters $A_1$($A_2$) representing the amplitude, $\mu_1$($\mu_2$) - the mean value, and $\sigma_1$($\sigma_2$) - the standard deviation of the first (second) Gaussian distribution are shown in Table \ref{table:parameters_guassian}.
The first Gaussian, depicted with a dotted line, most likely reflects the presence of the well-aligned brain fibers, because it is characterized by a low value of parameters $\mu_1$ (about $7^{\circ}-10^{\circ}$) and $\sigma_1$ (about $3^{\circ}-5^{\circ}$). 
The amplitude $A_1$ monotonically decreased with TTC increase: $A_1$(HT+WM) $>A_1$(IZ+WM) $>A_1$(TC+WM), thus, suggesting gradual loss of the aligned brain fibers in WM with tumor growth (Table \ref{table:parameters_guassian}).
The second Gaussian, depicted with a dashed line, most likely corresponds to regions with higher azimuth local variability related to brain fiber 1) crossing, 2) out of imaging plane orientation, or 3) perturbation due to tumor growth. 
It is characterized by the large values of parameter $\mu_2$ varying between $20^{\circ}-35^{\circ}$ and parameter $\sigma_2$ - between $13^{\circ}-15^{\circ}$ (Table \ref{table:parameters_guassian}).
The amplitude $A_2$ monotonously increased with TTC increase: $A_2$(HT+WM) $<A_2$(IZ+WM) $<A_2$(TC+WM), indicating that the proportion of regions with aligned fibers decreases with tumor growth.

\begin{table}[h!]
\centering
    \caption{Results of two-Gaussian fit.}

        \begin{tabular}{|c|c|c|c|c|c|c|}
            \hline
            &
            \textbf{$\mu_1$}&
            \textbf{$\sigma_1$}&
            \textbf{$A_1$}&
            \textbf{$\mu_2$}&
            \textbf{$\sigma_2$}&
            \textbf{$A_2$}\\ 
            \hline
            
            \textbf{HT}&
            7.02             &
            2.76             &
            0.73             &
            19.76            &
            13.46            &
            0.46\\
            \textbf{IZ} & 
            9.13             &
            4.17             &
            0.64             &
            28.02            &
            15.58            &
            0.85 \\ 
            \textbf{TC} & 
            11.55            & 
            4.39             & 
            0.14             &
            35.48            &
            13.12            &
            0.98  \\ \hline
        \end{tabular}
        
    \label{table:parameters_guassian}
\end{table}

In summary, we found differences in the distributions for depolarization and linear retardance for HBT and NBT,
The main finding is that the azimuth local variability was larger in the (NP+WM) regions compared to (HT+WM) regions, with noise distributions differing between HBT and NBT.

\section{Discussion}

\subsection{Polarimetric parameters in brain tumor specimens}

We report our newly developed neuropathology protocol, that enables to cross-reference polarimetric and histological data to assess HBT and NBT polarimetric properties.
This protocol allows for the generation of a reliable GT, enhancing the analysis of polarimetric properties of various brain tumor types.
An image processing pipeline was designed to ensure a highly accurate registration between histological and polarimetric images.
The protocol's design, particularly the H\&E staining of the first and last cryosections, ensured a direct correlation between histology and polarimetry for the two polarimetric measurements.
This comprehensive approach has also substantially increased the database size, as each tissue sample undergoes dual measurements, on different surfaces, due to the cryosectioning step.
Prior studies on polarimetric parameters in human organ tumor samples \cite{Vizet2017, Kupinski2018, Qi2023, Tata2016, Vahidnia2021} revealed a consistent reduction in depolarization values in NP regions, aligning with our findings of a slight decrease in depolarization values in both GM and WM.
Variations in the absolute value of change in depolarization across multiple studies arise from organ-specific microstructures and diverse methodologies.
Some studies also noted important drops in linear retardance values within NP areas, consistent with our results showing a marked decrease in linear retardance values in (NP+WM) regions of brain samples.
The reduction in linear retardance values was expected and observed in NBT in WM only, as fiber tracts are absent in GM.
Once again, the absolute value of change in the linear retardance varied across the different studies.

\subsection{Randomization of the azimuth of the optical axis}

Our study primarily focused on an increase in the azimuth local variability in (NP+WM) regions.
A similar pronounced azimuth randomization was observed in the previous polarimetric studies of uterine cervix \cite{Vizet2017}, where the values of standard deviation of the azimuth of the optical axis, computed for 3$\times$3 pixel packs, were notably higher in a polyp region compared to the surrounding HT cervical tissue.
The azimuth local variability in (TC+WM) regions for NBT exceeds the noise in the BG zones of the images while the distribution of the parameter is almost equivalent to that of the BG zone for (TC+GM) regions.
This discovery suggests that the origins of the azimuth local variability observed within (TC+WM) zones might be traced back to the distinct structural characteristics within the tumor tissue.
We assume that the presence and growth of tumors within the WM regions may perturb and eventually destroy aligned fibers in WM of brain, thus, leading to the increase in azimuth local variability level.
Understanding the specific reasons of this azimuth local variability increase is of paramount importance, as such insight has the potential to enhance the accuracy and reliability of the IMP in the detection and diagnosis of brain tumors.

We considered tissue microstructure alterations as a potential cause for azimuth randomization as tumors inherently disturb the normal architecture of surrounding HBT \cite{Quail2017}.
We identified two mechanisms potentially explaining the increased azimuth local variability in (NP+WM) brain tissue, likely occurring sequentially.
Initially, tumor growth leads to a disrupted arrangement of cells, extracellular matrix components, and blood vessels.
The physiological fiber orientation would consequently be perturbed due to tumor cell infiltration and/or proliferation.
At this stage, the WM would exhibit detectable linear retardance, while displaying high level of azimuth local variability, similar to the sample in Fig. \ref{fig:combined_results_tumor_90}c.
Secondly, a reduction, somewhat fragmentation of the nerve fibers may happen due to the physical pressure exerted by the tumor’s solid components.
Such pressure results in reduced local blood flow, thus, neuronal death with subsequent fiber degeneration \cite{Seano2019}.
At this stage, most of the studied (NP+WM) brain samples exhibited both low linear retardance and high azimuth local variability values.

\subsection{Implications for further development}
The identification of azimuth local variability as a potential polarimetric biomarker in NBT WM highlights key considerations for future ML algorithm development for accurate tumor boundaries delineation.
Recent studies successfully employed IMP and ML for NP tissue classification \cite{Luu2021, Robinson2023, Sampaio2023}.
The randomization of the azimuth of the optical axis in the (NP+WM) brain tissue that is linked to the increase in azimuth local variability may be explored by ML algorithms. These algorithms have the capability to learn and recognize subtle patterns or deviations in the spatial distribution of azimuth local variability, which might not be readily perceptible to the human eye.

The brain comprises two distinct tissue types, GM and WM, with the latter exhibiting moderate birefringence due to its unique microstructure, while the former lacks this property.
Consequently, tumor growth within each brain tissue type induces distinct effects on WM and GM polarimetric parameters.
The anticipated variations in polarimetric response to tumor growth between these brain tissue types highlight the influence of both TCC and tumor location (WM or GM) on the polarimetric parameters of brain tumor samples.

\section{Conclusion}

In this study, we have undertaken a comprehensive evaluation of polarimetric parameters in NBT, differentiating between HBT and NBT regions.
Our analysis reveals the drop in depolarization values in the GM and WM regions of NBT compared to HBT, and the drop in linear retardance values within the WM zones of NBT only.
Furthermore, the discovery of pronounced difference in the randomization of the azimuth of the optical axis within the WM zones of HBT and NBT indicates a valuable opportunity for the future development of ML algorithms for brain tumor delineation.
By creating a novel method for matching polarimetric dataset with histological data, we have established a robust foundation for further research in this area.
Taken together, this study represents an important step forward in our understanding of brain tumor optical characterization and opens an avenue to more accurate brain tumor diagnosis and treatment strategies.

\section*{Acknowledgment}
We would like to extend our heartfelt gratitude to the Translation Research Unit, Institute of Tissue Medicine and Pathology, University of Bern for their assistance and expertise in histological processing of brain tumor samples.

\section*{Data availability statement}
The codebase used to generate the results of the current study, as well as a few examples of polarimetric measurements of brain tumor are available in the following repository: \href{https://osf.io/rsxgp/?view\_only=c1cc83138a494606820e86c173d8d6f2}{https://osf.io/rsxgp}.

\newpage
\clearpage

\end{document}